\documentclass[a4paper]{article}
\usepackage[utf8]{inputenc}
\usepackage {amsmath, amssymb, mathrsfs, graphics, setspace}
\usepackage{braket}
\usepackage{multirow}
\usepackage{amsthm}
\usepackage{bbold}
\usepackage{cancel}
\usepackage{slashed}
\usepackage{color}
\usepackage{bbm}
\usepackage{amsfonts}
\usepackage{ytableau}
\usepackage{graphicx}
\usepackage{cite}
\usepackage{nicefrac}
\usepackage[]{authblk}
\usepackage{setspace}
\usepackage{placeins}
\usepackage{abstract}
\usepackage[linktocpage=true]{hyperref}
\usepackage{xcolor}
\hypersetup{
    colorlinks,
    linkcolor={blue!90!black},
    citecolor={blue!90!black},
    urlcolor={blue!90!black}
}

\newcommand{\idd}{\mathbb{1}}

\renewcommand{\vec}{\boldsymbol}    

\newcommand{\smatrix}{\begin{pmatrix}}
\newcommand{\cmatrix}{\end{pmatrix}}

\usepackage{mathtools}
\newlength{\myeqskip}  
\AtBeginDocument{%
    \setlength\abovedisplayskip{\myeqskip}%
    \setlength\belowdisplayskip{\myeqskip}%
    \setlength\abovedisplayshortskip{\myeqskip-\baselineskip}%
    \setlength\belowdisplayshortskip{\myeqskip}}

\begin{document}

\title{\vspace{-3.0cm} \fontsize{15}{18} \textbf{Bayesian solution to the inverse problem and its relation to Backus-Gilbert methods}\vspace{0.4cm}}

\author[a]{{Luigi Del Debbio}\thanks{\href{mailto:luigi.del.debbio@ed.ac.uk}{luigi.del.debbio@ed.ac.uk}}}
\author[b]{{Alessandro Lupo}\thanks{\href{mailto:alessandro.lupo@cpt.univ-mrs.fr}{alessandro.lupo@cpt.univ-mrs.fr}}}
\author[c]{{Marco Panero}\thanks{\href{mailto:marco.panero@unito.it}{marco.panero@unito.it}}}
\author[c]{{Nazario Tantalo}\thanks{\href{mailto:nazario.tantalo@roma2.infn.it}{nazario.tantalo@roma2.infn.it}}}

\affil[a]{Higgs Centre for Theoretical Physics, School of Physics \& Astronomy, The University of Edinburgh, Peter Guthrie Tait Road, Edinburgh EH9 3FD, United Kingdom\vspace{0.5cm}}

\affil[b]{Aix Marseille Univ, Université de Toulon, CNRS, CPT, Marseille, France\vspace{0.5cm}}

\affil[c]{Department of Physics, University of Turin \& INFN, Turin\\
Via Pietro Giuria 1, I-20125 Turin, Italy\vspace{0.5cm}}

\affil[d]{University and INFN of Roma Tor Vergata\\
	Via della Ricerca Scientifica 1, I-00133, Rome, Italy}
\date{}

{\let\newpage\relax\maketitle}

\begin{abstract}
\centering
\vspace{-0.3cm}
\begin{minipage}{\dimexpr\paperwidth-7cm}
The problem of obtaining spectral densities from lattice data has been receiving great attention due to its importance in our understanding of scattering processes in Quantum Field Theory, with applications both in the Standard Model and beyond. The problem is notoriously difficult as it amounts to performing an inverse Laplace transform, starting from a finite set of noisy data. Several strategies are now available to tackle this inverse problem. In this work, we discuss how Backus-Gilbert methods, in particular the variation introduced by some of the authors, relate to the solution based on Gaussian Processes. Both methods allow computing spectral densities smearing with a kernel, whose features depend on the detail of the algorithm. We will discuss such kernel, and show how Backus-Gilbert methods can be understood in a Bayesian fashion. As a consequence of this correspondence, we are able to interpret the algorithmic parameters of Backus-Gilbert methods as hyperparameters in the Bayesian language, which can be chosen by maximising a likelihood function. By performing a comparative study on lattice data, we show that, when both frameworks are set to compute the same quantity, the results are generally in agreement. Finally, we adopt a strategy to systematically validate both methodologies against pseudo-data, using covariance matrices measured from lattice simulations. In our setup, we find that the determination of the algorithmic parameters based on a stability analysis provides results that are, on average, more conservative than those based on the maximisation of a likelihood function.

\end{minipage}
\end{abstract}
\vspace{0.6cm}

\bigskip
\thispagestyle{empty} 

\section{Introduction}
The non-perturbative knowledge of spectral densities is of primary importance in the study of hadrons and their interactions. In lattice calculations, spectral densities can be accessed from Euclidean correlation functions by performing an inverse Laplace transform. The operation is difficult and requires delicate treatments. Nonetheless, increasing attention has been given to the problem, resulting in a growing literature of computational strategies~\cite{Asakawa:2000tr, Burnier:2013nla, Hansen:2019idp, Kades:2019wtd, Horak:2021syv, Bailas:2020qmv, Bergamaschi:2023xzx, Buzzicotti:2023qdv, Bruno:2024fqc}, and applications in different contexts of particle physics, such as the computation of scattering amplitudes~\cite{Bulava:2019kbi, Frezzotti:2023nun, Frezzotti:2024kqk, Patella:2024cto}, inclusive decays~\cite{Gambino:2020crt, Bulava:2021fre, Gambino:2022dvu,  ExtendedTwistedMassCollaborationETMC:2022sta, Evangelista:2023fmt, Barone:2023tbl, Alexandrou:2024gpl}, spectroscopy~\cite{DelDebbio:2022qgu, Bennett:2024cqv, Pawlowski:2022zhh, Panero:2023zdr} and QCD at finite temperature~\cite{Meyer:2007ic, Meyer:2007dy, Aarts:2007wj, Caron-Huot:2009ncn, Meyer:2011gj, Aarts:2011sm, Aarts:2012ka, Aarts:2013kaa, Aarts:2014nba, Ding:2015ona, Rothkopf:2019ipj, Itou:2020azb, Altenkort:2022yhb, Bonanno:2023ljc, Bonanno:2023thi, Aarts:2023vsf}. The difficulty stems from the fact that, due to the finite number of lattice data points and the presence of uncertainties on lattice correlators, the inversion of the Laplace transform is an ill-defined problem which needs to be regularised in order to yield a solution that is stable within the uncertainties on the lattice data. Moreover, due to the finite volume of the lattice, spectral densities are a sum of Dirac delta-functions. In order to manage them numerically, a smearing procedure is needed, which returns a smooth, well-behaved function. Once the spectral density is smeared, its infinite-volume limit can be studied in a mathematically well-defined way, and comparisons with experiments are possible, provided the continuum limit has also been taken. Smearing is a mandatory step of any lattice calculations of spectral densities.

The problem of numerical stability has been extensively studied, producing a variety of regularisations: Backus-Gilbert methods~\cite{10.1111/j.1365-246X.1968.tb00216.x}, in particular its formulation from Ref.~\cite{Hansen:2019idp}, Bayesian approaches~\cite{Horak:2021syv} and machine learning techniques~\cite{ Kades:2019wtd, Buzzicotti:2023qdv}. This work presents a comparison between Bayesian and Backus-Gilbert approaches, focusing on the relation between the solutions that are obtained in these two frameworks. In this sense, this work expands on the studies started in Refs.~\cite{10.1093/gji/ggz520, DelDebbio:2021whr, Candido:2023nnb}. Our contributions here are the following: we highlight an exact equivalence between the solution of the inverse problem obtained with Gaussian Processes (GP) using a specific class of priors, and the Backus-Gilbert procedure of Ref.~\cite{Hansen:2019idp} (HLT in short) which allows controlling the smearing kernel of the solution. Moreover, we show how the algorithmic parameters of Backus-Gilbert methods can be interpreted as hyperparameters of a prior, and therefore chosen accordingly by studying likelihood functions. The choice of Backus-Gilbert parameters is a delicate issue, highlighting the importance of this correspondence. Finally, we give a detailed discussion on the relation among the systematics of the two approaches, based on a series of tests against both mock and real data. The software used in this work to solve the inverse problem can be found at~\cite{Forzano:2024}.

\section{Formulation of the problem}
We are concerned with the calculation of the infinite-volume spectral density $\rho(E)$, from the correlation function of two gauge-invariant operators separated by a Euclidean time t:
\begin{equation}
    C_{LT}(t) = \int_{0}^\infty dE \; b_T(t,E) \, \rho_{LT}(E) \; ,
\end{equation}
which is obtained from Monte Carlo simulations of the theory on a lattice with  spacing $a$, finite volume $L^3$, and temporal extent $T$. In the limit of infinite time extent, the correlator is related to the spectral density by a Laplace transform,
\begin{equation}\label{eq:laplace_transform}
    C_L(t) = \lim_{T \rightarrow \infty} C_{TL}(t) = \int_{0}^\infty dE \; e^{-tE} \rho_L(E) \; ,
\end{equation}
There are several obstacles, in our context, to the inversion of Eq.~\eqref{eq:laplace_transform}. In a quantum field theory, spectral densities are defined as tempered distributions, which become especially unmanageable in a finite volume, where they are sums of Dirac $\delta$-functions. For this reason, it is necessary to introduce a Schwartz function $\mathcal{S}_\sigma(\omega-E)$ and consider the smeared spectral density, 
\begin{equation}\label{eq:smearing}
    \rho_L(\sigma;\omega) = \int dE \, \mathcal{S}_\sigma(\omega-E) \, \rho_L(E) \, ,
\end{equation}
which is a regular function even at a finite $L$. For convenience, the function $\mathcal{S}_\sigma$ is parametrised by a smearing radius $\sigma$, so that in the limit of $\sigma \rightarrow 0$ we recover a $\delta$-function,
\begin{equation}
    \lim_{\sigma \rightarrow 0} \mathcal{S}_\sigma(\omega-E) = \delta(\omega-E) \; .
\end{equation}
In this way, the infinite-volume limit of the spectral density can be defined as
\begin{equation}
    \rho(\omega) = \lim_{\sigma \rightarrow 0} \lim_{L \rightarrow \infty} \rho_{L}(\sigma;\omega) \; .
\end{equation}
The task then shifts to the computation of $\rho_{L}(\sigma;\omega)$, yet difficulties persist. While the smeared spectral density has support over a continuous set of energies, the correlator $C(t)$ is only known at a finite set of points,
\begin{equation}
    0 < \tau \leq \tau_{\max} \; , \;\;\; \;\; \tau = t / a \; ,
\end{equation}
limiting the amount of information we are capable of extracting. The other major obstruction, which makes our problem ill-defined, is the finite precision of the lattice data $C(t)$. If we could compute the correlator without any uncertainty for infinitely many discrete values of the Euclidean time, $C^{\rm exact}$, the related smeared spectral density can be expressed as a linear combination of such correlators,
\begin{equation}\label{eq:rho_equals_sum_gt_ct_EXACT}
    \rho^{\rm exact}(\sigma; \omega) = \sum_{\tau=1}^{\infty} g^{\rm exact}_\tau(\sigma;\omega) \, C^{\rm exact}(a \tau) \; .
\end{equation}
Even with a finite number of data points, $
\tau_{\rm max}$, the approximation 
\begin{equation}\label{eq:rho_equals_sum_gt_ct_FINITE}
    \rho(\sigma ; \omega) = \sum_{\tau =1}^{\tau_{\rm max}} g_\tau(\sigma; \omega) \, C(a \tau) \; ,
\end{equation}
is unambiguously defined. Eqs.~\eqref{eq:rho_equals_sum_gt_ct_EXACT} and \eqref{eq:rho_equals_sum_gt_ct_FINITE} are, however, the result of highly fine-tuned cancellations between individual terms~\cite{Hansen:2019idp, Buzzicotti:2023qdv}. These are not possible when the correlators are noisy. A regularisation must be provided, so that the solution is stable within the uncertainty on $C(a \tau)$.

In this work, we intend to clarify how two common solutions to the inverse problem, one relying on the Backus-Gilbert regularisation of Ref.~\cite{Hansen:2019idp}, the other on stochastic processes, deal with the aforementioned issues. We will show how both approaches satisfy the conditions that are necessary to treat the problem, i.e. how they both provide a smeared solution that is stable with respect to the statistical fluctuations of the input correlators. We begin by noticing that any solution that is expressed as a linear combination of the correlators, as in Eq.~\eqref{eq:rho_equals_sum_gt_ct_FINITE}, yields a smeared spectral density,
\begin{equation}\label{eq:every_reconstruction_is_smeared}
   \sum_{\tau =1}^{\tau_{\rm max}} g_\tau(\sigma;\omega) \, C(a \tau) =   \int dE \left( \sum_{\tau =1}^{\tau_{\rm max}} g_\tau(\sigma;\omega) \,  b_T(a \tau,E)  \right)\, \rho_L(E) \; ,
\end{equation}
with a smearing kernel 
\begin{equation}
    \mathcal{S}_\sigma(E,\omega) = \sum_{\tau =1}^{\tau_{\rm max}} g_\tau(\sigma;\omega) \,  b_T(a \tau,E)  \; .
\end{equation}
The approaches we describe in this paper are based on inherently different assumptions, which lead to different computational strategies to determine the coefficients $g_\tau$.

\section{Gaussian Processes}\label{sec:GP}
In the context of Bayesian inference with Gaussian Processes, the problem is formulated in a probabilistic fashion, so that one can seek a probability distribution over a functional space of spectral densities, rather than the spectral density itself. As it will be shown, such probability distribution will take into account the data uncertainty and, if available, prior knowledge about the solution. This approach is especially attractive due to the possibility of obtaining analytic expressions for the predictions.

We represent the spectral density as a  stochastic field $\mathcal{R}(\omega)$, which is described by a GP centred around $\rho^{\rm prior}(\omega)$ and with covariance $\mathcal{K}^{\rm prior}(\omega,E)$. The prior probability measure is therefore
\begin{equation}
    \Pi[\mathcal{R}]  = \frac{1}{\mathcal{N}} \, \exp \biggr( -\frac{1}{2} \left| \mathcal{R} - \rho^{\rm prior} \right|^2_{\mathcal{K}^{\rm prior}} \biggr) \; , 
\end{equation}
where
\begin{equation}
    \left| \mathcal{R} - \right.
    \left.\rho^{\rm prior} \right|^2_{\mathcal{K}^{\rm prior}} =    \int dE_1 \int dE_2 
    \left[ \mathcal{R}(E_1) - \rho^{\rm prior}(E_1) \right] \mathcal{K}_{\rm prior}^{-1}(E_1,E_2)
     \left[\mathcal{R}(E_2) - \rho^{\rm prior}(E_2) \right]
\end{equation}
and the normalisation
\begin{equation}
    \mathcal{N} = \int \mathcal{D} \mathcal{R} \; \Pi[\mathcal{R}] \; .
\end{equation}
In the previous expressions, $\mathcal{D} R$ represents the functional integration measure over the field variable $\mathcal{R}$. 
By definition, the expectation value for $\mathcal{R}(E)$ from the prior distribution is
\begin{equation}
    \rho^{\rm prior}(\omega) = \int \mathcal{D}\mathcal{R}\, \Pi[\mathcal{R}] \; \mathcal{R}(\omega)  \; .
\end{equation}
We then introduce the noise $\vec{\eta} \in \mathbb{R}^{\tau_{\rm max}}$ which takes into account the uncertainty in the lattice data and is represented as a real-valued stochastic variable, for which we assume a multivariate Gaussian distribution with zero mean, 
\begin{equation}\label{eq:eta_gauss_noise}
    \mathbb{G}[\vec{\eta}, \text{Cov}_d] = \frac{1}{\sqrt{\text{det} (2\pi \text{Cov}_d)}} \exp \left( -\frac{1}{2}  \vec{\eta}\; \text{Cov}_d^{-1} \;  \vec{\eta}  \right) \, ,
\end{equation}
$\text{Cov}_d$ being the covariance matrix of the correlators. We can now introduce the stochastic variable associated to the correlator, $\vec{{\mathcal{C}}} \in \mathbb{R}^{\tau_{\rm max}}$ with entries:
\begin{equation}\label{eq:laplace_transform_with_noise}
    \mathcal{C}(t) = \int dE \, b_T(t,E) \, \mathcal{R}(E) + \eta(t) \; .
\end{equation} 
Given the distributions of Eq.~\eqref{eq:laplace_transform_with_noise} and~\eqref{eq:eta_gauss_noise} we can evaluate the covariance associated to the variable $\vec{\mathcal{C}}$. It is straightforward to show that
\begin{equation}\label{eq:CCSigmaHatPlusB}
    \begin{split}
        \braket{\mathcal{C}(t_1)\mathcal{C}(t_2)} & = \int d\vec{\eta}\;\mathcal{D}\mathcal{R}\; \mathcal{C}(t_1) \, \mathcal{C}(t_2) \;
        \mathbb{G}[\vec{\eta}, \text{Cov}_d] \,
        \Pi[\mathcal{R}] \,  \\[8pt]
        & = \Sigma_{t_1t_2} + \left(\text{Cov}_d\right)_{t_1t_2} \; ,
    \end{split}
\end{equation}
where we defined
\begin{equation}\label{eq:SigmaDef}
    \Sigma_{t_1t_2} = \int dE_1 \int \; dE_2 \; b_T(t_1,E_1) \, \mathcal{K}^{\rm prior}(E_1,E_2) \, b_T(t_2,E_2) 
\end{equation}

In order to predict the value of the spectral density at the energy $\omega$, we need to extend the dimensionality of the covariance of Eq.~\eqref{eq:CCSigmaHatPlusB} to include the indirect observation of $\rho(\omega)$. To this end, we introduce the vector $\vec{F} \in \mathbb{R}^{\tau_{\rm max}}$ of components
\begin{equation}\label{eq:Fvector}
    F_t(\omega) = 
    \langle \mathcal{C}(t) \mathcal{R}(\omega)\rangle
    =
    \int d\vec{\eta}\, 
    \mathcal{D}\mathcal{R} \; \mathcal{C}(t) \, \mathcal{R}(\omega)\,  
    \mathbb{G}[\vec{\eta}, \text{Cov}_d] \,\Pi[R]\;   , 
\end{equation}
together with the scalar $F_*$,
\begin{equation}\label{eq:f_star}
    F_*(\omega) = \int \mathcal{D}\mathcal{R} \; \mathcal{R}(\omega)^2
    \; \Pi[\mathcal{R}]  = \mathcal{K}^{\rm prior}(\omega,\omega) \; .
\end{equation}
The total covariance, which has the additional information at the energy $\omega$, is then given by 
\begin{equation}\label{eq:total_covariance}
    \Sigma^{\rm tot} = \begin{pmatrix} F_*(\omega) & \vec{F}(\omega)^T \\ \vec{F}(\omega) & \Sigma+\text{Cov}_d \; .
    \end{pmatrix}
\end{equation}
Let $C^{\rm obs}(t)$ be the correlator measured on the lattice by averaging over a gauge ensemble. We further denote $\vec{C}^{\, \rm prior}$ as the vector $\vec{\mathcal{C}}$ evaluated at $\mathcal{R} = \rho^{\rm prior}$, whose components are:
\begin{equation}
     \mathcal{C}(t) \, |_{\rho^{\rm prior}} = C^{\rm prior}(t) = \int dE \, b_T(t,E) \, \rho^{\rm prior}(E) \; .
\end{equation}
The joint probability density for $\mathcal{R}(\omega)$ and $\mathcal{C}(t)$, which is under our assumption the Gaussian
\begin{equation}
    \mathbb{G} \left[\mathcal{R} - \rho^{\rm prior} , \, \vec{\mathcal{C}} - \vec{C}^{\; \rm prior} ; \, \Sigma^{\rm tot} \right] \; ,
\end{equation}
can be factorised, as shown in Appendix~\ref{app:sec:LDU}, in the product between the posterior probability density for $\mathcal{R}(\omega)$ given its prior distribution and set of measurements for the correlator, and the likelihood of the data. These are, in this setup, both Gaussian:
\begin{equation}\label{eq:conditional_probability_factorised}
    \mathbb{G} \left[\mathcal{R} - \rho^{\rm prior} , \, \vec{\mathcal{C}} - \vec{C}^{\; \rm prior} ; \, \Sigma^{\rm tot} \right] = \mathbb{G} \left[ \mathcal{R}- \rho^{\rm post}; \, \mathcal{K}^{\rm post}\right]  \; \mathbb{G} \left[ \vec{\mathcal{C}} - \vec{C}^{\; \rm prior} ; \, \Sigma + \text{Cov}_d \right] \; .
\end{equation}
The posterior Gaussian distribution for the spectral density is centred around:
\begin{equation}\label{eq:naive_gp_mean}
    \left. \rho^{\rm post}(\omega) \right|_{\mathcal{C}=C^{\rm obs}} = \rho^{\rm prior}(\omega) + \vec{F}^T(\omega)  \frac{1}{\Sigma + \text{Cov}_d}   \left( \vec{C}^{\;\rm obs} - \vec{C}^{\; \rm prior} \right) \; ,
\end{equation}
and has variance
\begin{equation}\label{eq:naive_gp_variance}
    \left. \mathcal{K}^{\rm post}(\omega,\omega)\right|_{\mathcal{C}=C^{\rm obs}} =
     \mathcal{K}^{\rm prior}(\omega,\omega) - \vec{F}^T(\omega) \frac{1}{\Sigma + \text{Cov}_d}  \vec{F}(\omega) \; .
\end{equation}
In order to make contact with Eq.~\eqref{eq:rho_equals_sum_gt_ct_FINITE}, we introduce the coefficients
\begin{equation}\label{eq:gt_naive_GP}
    \vec{g}^{\rm \, GP} (\omega) =  \vec{F}^T  \frac{1}{\Sigma + \text{Cov}_d} \; .
\end{equation}
We now pause to make some comments. First, the previous equations clarify how the problem is regularised by formulating it in terms of probability distributions. The symptom of numerical instability, which leads to the very large coefficients of Eq.~\eqref{eq:rho_equals_sum_gt_ct_EXACT}, is in fact the condition number of the model covariance $\Sigma$ of Eq.~\eqref{eq:SigmaDef}, which grows exponentially for acceptable choices of the model covariance $\mathcal{K}$. In absence of error on the data, the magnitude of the coefficients are uniquely determined by the inverse of  the matrix $\Sigma$, which is very large. For noisy data, the covariance $\text{Cov}_d$ is added to the matrix $\Sigma$, providing a cut-off for its lower modes. The magnitude of the coefficients is now determined by the conditioning of $\Sigma + \text{Cov}_d$, which depend on the prior $\mathcal{K}^{\rm prior}$ and the size of the error on the data. We will later show how the choice of $\mathcal{K}^{\rm prior}$ can be used to tune the cut-off on the low modes of $\Sigma$.

Another important remark concerns the smearing kernel that one has implicitly introduced in Eq.~\eqref{eq:naive_gp_mean}. This is given by
\begin{equation}\label{eq:GP_smearing_kernel}
    \mathcal{S}^{\rm GP}(\omega,E) = \sum_{\tau=1}^{\tau_{\rm max}} \vec{g}^{\rm \, GP} (\omega) \,b_T(a\tau ,E) \; .
\end{equation}
We stress that the centre of the posterior distribution, $\rho^{\rm post}(\omega)$ is an estimate for a spectral density that is smeared with $\mathcal{S}^{\rm GP}$. The unsmeared spectral density can be obtained if the prior is engineered such that $\mathcal{S}^{\rm GP}(\omega-E)$ is a Dirac-$\delta$. Such a distribution cannot be obtained by a finite linear combination of the regular functions $b_T(t,E)$, and one would need to extrapolate the result at vanishing smearing radius.

We conclude this section with a crucial remark on the nature of the prior. The functional behaviour of the prediction is related to the prior through Eqs.~\eqref{eq:naive_gp_mean} and~\eqref{eq:naive_gp_variance}. In particular, the covariance of the prior is known to affect the typical correlation length of the posterior. On the lattice finite-volume spectral densities are targeted, therefore this length should be smaller than the typical spacing between energy levels in order to capture the features of the underlying physics.

\subsection{Choice of the priors}
In this work we consider, for the prior model covariance, variations around a Gaussian $G_\epsilon(\omega-E)$:
\begin{equation}\label{eq:Kprior_gaussian}
    \mathcal{K}(\omega,E) = \frac{e^{\alpha E}}{\lambda}\frac{e^{-\frac{(\omega-E)^2}{2\epsilon^2}}}{\sqrt{2\pi} \epsilon} \equiv \frac{e^{\alpha E}}{\lambda} G_\epsilon(\omega-E) \; ,
\end{equation}
where $\epsilon$, $\alpha$ and $\lambda$ are the defining parameters of the prior, which are often referred to as “hyperparameters” in the literature of GPs. The motivation behind the choice made in Eq.~\eqref{eq:Kprior_gaussian} are the following. First, a Gaussian is a common choice in the literature~\cite{Horak:2021syv,10.1093/gji/ggz520}, with width $\epsilon$ and amplitude $\lambda^{-1}$ as parameters to be chosen. In addition, the Gaussian allows to control the limit in which the covariance becomes diagonal by changing the parameter $\epsilon$. The term $e^{\alpha E}$ allows controlling deviations from the Gaussian case, and it will be a point of contact with the method of Ref.~\cite{Hansen:2019idp}. In order to set the notation, we specialise the previous equations to this choice of the prior:
\begin{equation}\label{eq:SigmaMat_GP_specialised}
    \frac{\Sigma^\epsilon_{tr}}{\lambda} = \int dE_1 \int dE_2\; b_T(E_1,t) \,  b_T(E_2,r)\, e^{\alpha E_1} \, \frac{G_\epsilon(E_1-E_2)}{\lambda} \; ,
\end{equation}
\begin{equation}\label{eq:Fvector_GP_specialised}
    \frac{F^{\epsilon}_t(\omega)}{\lambda} = \int dE \, b_T(t,E) \, e^{\alpha E} \; \frac{G_\epsilon(E-\omega)}{\lambda} \; ,
\end{equation}
where we chose to factorise $\lambda$ from the definitions. The resulting expression for the coefficients of Eq.~\eqref{eq:gt_naive_GP} is
\begin{equation}\label{eq:gt_naiveGP}
\vec{g}^{\rm GP}(\epsilon; \omega) = \vec{F}^{\, T}(\omega) \, \frac{1}{\Sigma^\epsilon + \lambda \, \text{Cov}_d}  \; .
\end{equation}
It is clear that $\lambda$ is parametrising the cut-off on the low modes of $\Sigma^\epsilon$, thus introducing a bias. When $\lambda$ becomes larger the role of the regularising term, $\text{Cov}_d$, is enhanced, and the coefficients $g_\tau$ become increasingly smaller. At smaller values of $\lambda$, the bias decreases, but it cannot be eliminated. Its dependence must be therefore addressed, its effect quantified and controlled. The role of such parameter is extensively discussed in the context of Backus-Gilbert methods, and in particular in its formulation of Ref.~\cite{Hansen:2019idp}, where it is usually prescribed to choose $\lambda$ such that the prediction for the spectral density is stable within statistical noise~\cite{Bulava:2021fre,DelDebbio:2022qgu,ExtendedTwistedMassCollaborationETMC:2022sta} upon variations of $\lambda$. From this ``stability analysis'', the bias is assumed to be absorbed into the statistical error, a procedure that has been validated numerically, see for instance Refs.~\cite{Bulava:2021fre, ExtendedTwistedMassCollaborationETMC:2022sta, Bennett:2024cqv}. In the context of GPs, on the other hand, the hyperparameters (including $\lambda$) are selected so that the resulting probability of observing the data,
\begin{equation}
\mathbb{G} \left[ \vec{C}^{\, \rm obs} - \vec{C}^{\; \rm prior} ; \, \Sigma + \text{Cov}_d \right] \; , 
\end{equation}
is maximised~\cite{Horak:2021syv,10.1111/j.1365-246X.1968.tb00216.x}. Equivalently, one minimises the "negative logarithmic likelihood" (NLL)
\begin{multline}\label{eq:NLL}
\frac{\tau_{\rm max}}{2}\, \text{Log} (2\pi) +\frac{1}{2} \, \text{Log}\, \text{det} \left( \Sigma + \text{Cov}_d \right) +\frac{1}{2} \, (\vec{C}^{\, \rm obs}- \vec{C}^{\; \rm prior}) \frac{1}{\Sigma+\text{Cov}_d}  (\vec{C}^{\, \rm obs}- \vec{C}^{\; \rm prior}) \; .
\end{multline}

\begin{figure}[tbh]
    \centering
    \includegraphics[width=0.49\textwidth]{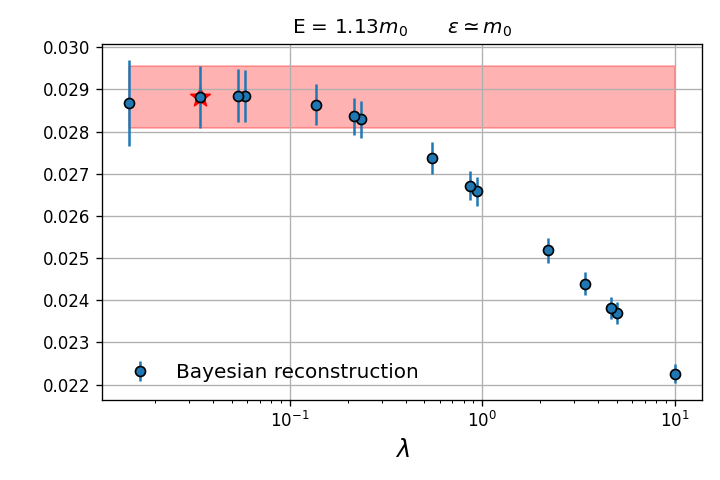}
    \includegraphics[width=0.49\textwidth]{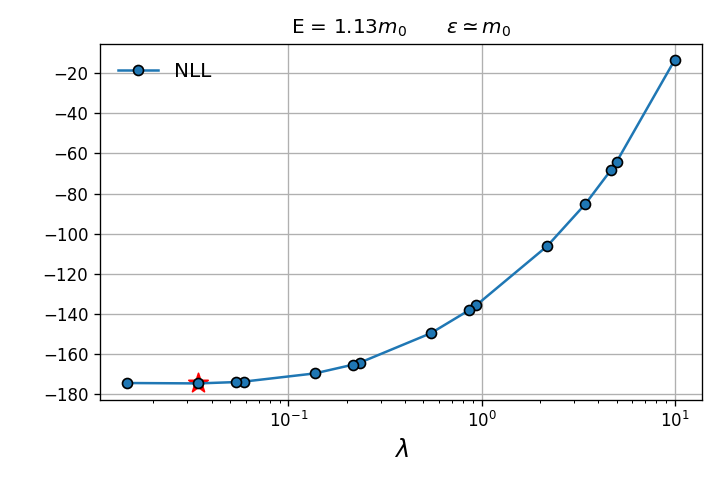}
    \caption{Combination of the stability analysis used (left) and a scan of the NLL (right). The smeared density shown in the left panel is obtained at a specific energy, and is evaluated from GPs, i.e. from Eqs.~\eqref{eq:naive_gp_mean} and~\eqref{eq:naive_gp_variance} for the central value and the error respectively. The prior is the modified Gaussian of Eq.~\eqref{eq:Kprior_gaussian}, with $\epsilon \simeq m_0$, the latter being the ground state of the channel. We only display value corresponding to $\alpha=0$ because these correspond to a systematically smaller NLL. The purpose of this figure is to show that the treatment of the parameter $\lambda$ from Ref.~\cite{Hansen:2019idp} and GPs can lead to compatible results. Details about the lattice data used for this example are found in the main text.}
    \label{fig:naiveGPscan}
\end{figure}

In this setting, the fate of the bias introduced by $\lambda$ could be considered opaque. We therefore assess whether porting the ``stability analysis'' that is carried in Ref.~\cite{Hansen:2019idp} into the Bayesian setup, can add insights in this regard. In Fig.~\ref{fig:naiveGPscan}, we show how the choice of $\lambda$ affects the spectral reconstruction (left panel)\footnote{The result shown in Fig.~\ref{fig:naiveGPscan} is obtained from a pseudoscalar correlator of fermions in a higher representation, computed within the ensemble B3 generated by the authors in Ref.~\cite{DelDebbio:2022qgu}.} at a specific energy. The corresponding values of the NLL are shown in the right panel of the same figure, with the minimum value highlighted. For large values of $\lambda$, the smeared spectral density changes considerably, showing a large dependence on the prior. The corresponding values of the NLL are large. Remarkably, as the NLL approaches its minimum, the dependence on the prior softens. The horizontal band in the left panel of Fig.~\ref{fig:naiveGPscan} is obtained at the value of $\lambda$ that minimises the NLL, which is flagged by a star in the right panel. This result suggests that the treatment of the bias might not be drastically different in these two cases, despite the very different approaches.

\begin{figure}[tbh!]
    \centering
    \includegraphics[width=0.49\textwidth]{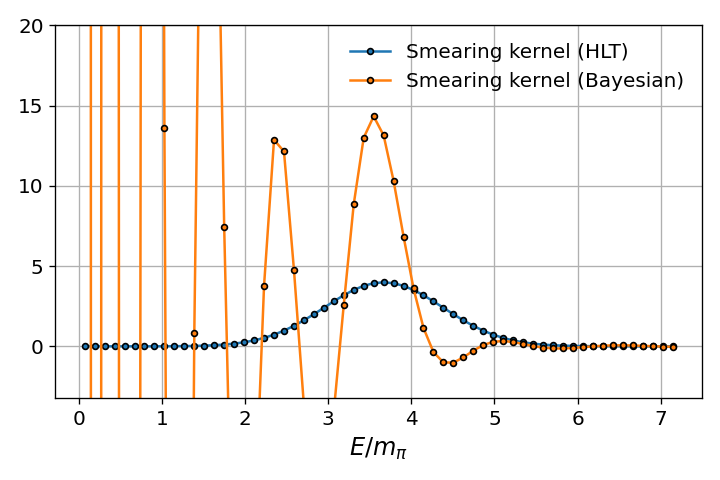}
    \includegraphics[width=0.49\textwidth]{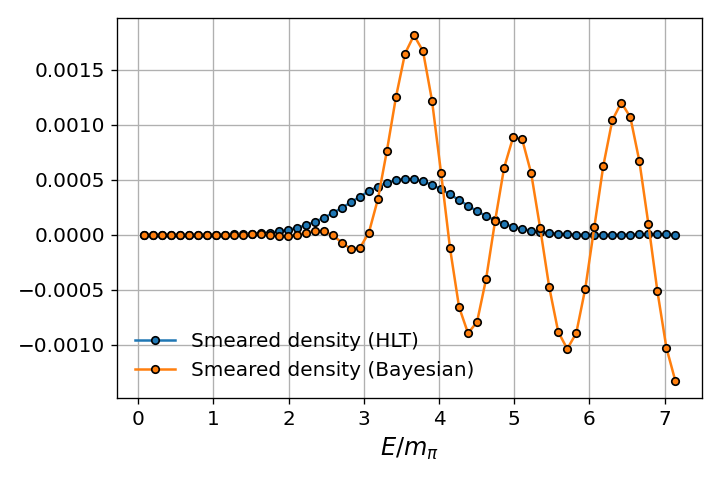}
    \caption{Left panel: examples of the function smearing the spectral density at the energy $E = 3.7 m_\pi$ in the Bayesian setup (orange) and from the HLT procedure (blue) using exact data. The latter targets a Gaussian kernel with a width of approximately $0.75$ in units of $m_\pi$, which is reconstructed with great precision given the lack of uncertainties on the input data. For the Bayesian calculation, we use the same Gaussian function as a prior, but we have less control on the output function, which in this case features oscillations with a length scale determined by the prior. The right panel displays the reconstructed smeared spectral densities from the same data. This example uses $t_{\rm max}=32$ data points.}
    \label{fig:naive_smearing_kernel}
\end{figure}

Another important remark, already stated in the previous Section, is that the value obtained at the minimum of the NLL (horizontal band in Fig.~\ref{fig:naiveGPscan}) has to be interpreted and understood as a spectral density that is smeared with the appropriate smearing kernel, which is not known a priori in the Bayesian formulation given in this section. In this respect, this method is closer to the original proposal of Backus and Gilbert rather than the modification of Ref.~\cite{Hansen:2019idp}. For this reason we find it instructive to show, in the left panel of Figure~\ref{fig:naive_smearing_kernel}, an example of the smearing kernel at a specific energy ($\omega= 3.7 m_\pi$) obtained with the Bayesian setup (orange), as opposed to the smearing kernel obtained from the method (in blue) of Ref.~\cite{Hansen:2019idp} (HLT in short) that will be described in the next Section. The Bayesian kernel is not known a priori: its behaviour is constrained by the covariance of the prior, which can produce a smoother or more rapidly changing function. In the example of Fig.~\ref{fig:naive_smearing_kernel} the prior is defined in Eq.~\ref{eq:Kprior_gaussian}, with $\epsilon= 0.75 m_\pi$. The output thus features, around $\omega= 3.7 m_\pi$, oscillation that have roughly a wavelength of $\epsilon$. In the right panel of the same figure, the smeared spectral density obtained with the two methods is also displayed. In this example, we have used synthetic data without any statistical noise, in order to showcase what each method does in the ideal limit of exact data. The input correlator contains a single state with $E/m_\pi = 3.7$.

\FloatBarrier
\section{Backus-Gilbert methods}\label{sec:hlt}

While methods derived from the seminal work of Backus and Gilbert~\cite{10.1111/j.1365-246X.1968.tb00216.x} do not arise from a probabilistic framework, the way the problem is regularised is numerically equivalent. In this work, we adopt HLT method~\cite{Hansen:2019idp}, which has some appealing features from the perspective of lattice simulations. The idea, which we briefly review, is to obtain a spectral density smeared with a chosen kernel, for instance the Gaussian $G_\sigma(\omega-E) = \exp(-(\omega-E)^2/2\sigma^2) / \sqrt{2\pi}\sigma$. Since the smearing function is identified in Eq.~\eqref{eq:every_reconstruction_is_smeared}, the idea of Ref.~\cite{Hansen:2019idp} is to define the coefficients entering Eq.~\eqref{eq:rho_equals_sum_gt_ct_FINITE} as those leading to the desired kernel,
\begin{equation}
     \sum_{\tau=1}^{\infty} \, g_\tau(\sigma;\omega) \, b_T(a\tau,E) = G_\sigma(\omega,E) \; .
\end{equation}
Since we only have a finite number of data, $\tau_{\rm max}$, we find an approximation for the coefficients in the previous equation by minimising the following functional:
\begin{equation}\label{eq:A_funcitonal}
    A[g(\omega)] = \int_{0}^\infty dE \, e^{\alpha E} \left| \sum_{\tau=1}^{\tau_{\rm max}} \, g_\tau(\sigma;\omega) \, b_T(a\tau,E) - G_\sigma(\omega-E) \right|^2 \; .
\end{equation}
The parameter $\alpha<2$ defines a class of norms that can be used to measure the distance from the desired kernel. Since Eq.~\eqref{eq:A_funcitonal} itself does not provide a stable minimum, one minimises the regularised functional
\begin{equation}\label{eq:W_functional}
    (1-\lambda') A[g] + \lambda' B[g] \; ,
\end{equation}
where 
\begin{equation}
    B[g] = \vec{g} \, \text{Cov}_d \, \vec{g} \; ,
\end{equation}
and $\lambda' \in (0,1)$ parametrises the extent of the regularisation. The solution that one obtains by minimising Eq.~\eqref{eq:W_functional} has a similar expression to Eq.~\eqref{eq:gt_naive_GP}, obtained in the context of GPs,
\begin{equation}\label{eq:gt_hlt}
    \vec{g}(\omega) = \vec{F}^{\sigma}(\omega)^T \, \frac{1}{ \Sigma^0 + \lambda \, \text{Cov}_d } \; ,
\end{equation}
where
\begin{align}\label{eq:gt_hlt_ingredients}
    & \Sigma^0_{tr} = \lim_{\epsilon \rightarrow 0} \Sigma^\epsilon_{tr} =  \int dE \, e^{\alpha E} \, e^{-(t+r)E} \; , \\[8pt]
    & F^\sigma_t(\omega)  = \int dE \, e^{\alpha E} \, G_\sigma(\omega-E) \, e^{-tE} \; ,  \\[8pt]
    & \lambda = \frac{\lambda'}{1-\lambda'} \in (0,\infty) \;.
\end{align}
The equation for the coefficients, \eqref{eq:gt_hlt}, is strikingly similar to the one arising in the context of GPs with a Gaussian model covariance. The matrix $\Sigma_0$ corresponds to the matrix $\Sigma^\epsilon$, from Eq.~\eqref{eq:SigmaMat_GP_specialised}, in the limit in which the covariance $\mathcal{K}^{\rm prior}$ becomes diagonal. The expression for the vector $\vec{F}$ is the same, with the difference that the parameters $\sigma$ in Eq.~\eqref{eq:gt_hlt_ingredients} represents the radius of the smearing Gaussian, while in Eq.~\eqref{eq:Fvector_GP_specialised} the parameter $\epsilon$ it is the radius of the model prior, which is again a Gaussian. A direct consequence is that the limit in which $\mathcal{K}^{\rm prior}$ is diagonal for the GP provides the very same solution of Ref.~\cite{Hansen:2019idp}, the latter taken in the limit of vanishing smearing radius. As stated above, this limit can only be extrapolated, since the inverse problem cannot be solved for a non-smeared spectral density.

Another parallelism concerns the fate of the parameter $\lambda$ which appears in Eqs~\eqref{eq:gt_hlt} and~\eqref{eq:gt_naive_GP} and was partially described in the previous section. Its role in regularising the problem is the same within GP and BG methods, see Eqs.~\eqref{eq:gt_naiveGP} and~\eqref{eq:gt_hlt}. This parameter is known in the literature of Backus-Gilbert methods as a trade-off parameter, determining the relative importance that we give to the minimisation of $A[g]$ or $B[g]$ in Eq.~\eqref{eq:W_functional}, and as stated before, it introduces a bias. As $\lambda$ approaches zero this bias, in the form of the systematic error due to approximate reconstruction of the smearing kernel
decreases, but the problem becomes numerically unstable. This is compensated by an increasing in the statistical error.  Conversely, as $\lambda$ increases, the solution becomes more stable, the statistical error decreases, but the bias dominates. The choice of $\lambda$ for a given set of data is a delicate problem. The smeared spectral density cannot depend on unphysical, algorithmic parameters.

As pointed out in Ref.~\cite{Bulava:2021fre} and as we recalled in the previous section, approaching smaller values of $\lambda$ there is a “stability region”, where the solution only fluctuates within the statistical errors when unphysical parameters vary, similarly to what happened in Fig.~\ref{fig:naiveGPscan}. While it is understood that the bias due to $\lambda$ cannot be fully removed, this analysis suggests that its effect can be absorbed within statistical noise. As an additional precaution, once the stability analysis provides a value $\lambda^*$, we repeat the calculation at a smaller value, e.g. $\lambda^* / 10$. If the difference is significant, it is accounted for as a systematic error.

\section{Bayesian inference for a smeared spectral density}
\label{sec:gphlt}
In this section, we show that Backus-Gilbert methods can be formulated in a probabilistic fashion, in terms of GPs. To our knowledge, this connection was first pointed out in Ref.~\cite{10.1093/gji/ggz520} and later, independently, in Ref.~\cite{DelDebbio:2021whr}. Again we will use the HLT formulation.~\cite{Hansen:2019idp}, but the idea can be extended.

In the HLT method a solution to the inverse problem is provided in terms of a spectral density that is smeared with a chosen kernel. Such solution cannot be directly compared with the one of Eq~\eqref{eq:naive_gp_mean}, unless the smearing function is the same in both cases. In order to directly compare the predictions, we therefore arrange the GP to predict the spectral density smeared with a chosen kernel, instead of letting the smearing as an implicit step. To this end, we associate a stochastic variable to the smeared spectral density,
\begin{equation}
    \mathcal{R}_\sigma(\omega) = \int dE \; \mathcal{S}_\sigma(\omega,E) \, \mathcal{R}(E) \; ,
\end{equation}
as opposed to the variable $\mathcal{R}(E)$ of Section~\ref{sec:GP}.  Repeating the derivation of Section~\ref{sec:GP} one lands on the following total covariance
\begin{equation}
\Sigma^{\rm tot} = \begin{pmatrix}
F^* & \vec{F} \\ \vec{F} & \Sigma + \text{Cov}_d
\end{pmatrix} \; ,
\end{equation}
where $\Sigma$ is defined in Eq.~\eqref{eq:SigmaDef}, while the other terms are now different:
\begin{equation}
F_*(\omega) = \int dE_1 \int dE_2 \; \mathcal{S}_\sigma(\omega,E_1) \, \mathcal{K}^{\rm prior}(E_1, E_2) \, \mathcal{S}_\sigma(E_2, \omega) \; ,
\end{equation}
\begin{equation}
F^{\sigma}_{t}(\omega) = \int dE_1 \int dE_2 \;  \, b_T(a\tau, E_1) \, \mathcal{K}^{\rm prior}(E_1, E_2) \, \mathcal{S}_\sigma(E_2,\omega) \, , 
\end{equation}
The important step towards matching with Ref.~\cite{Hansen:2019idp} is now to choose the model covariance to be diagonal,
\begin{equation}\label{eq:hltgp_prior}
\mathcal{K}^{\rm prior}(E_1, E_2) = \frac{e^{\alpha E}}{\lambda} \delta(E_1-E_2) \; .
\end{equation}
Beyond the idea of matching to HLT, this choice for the covariance of the prior is in fact the correct one when we account for the distributional nature of the underlying spectral density. The target function must be in fact allowed to reproduce the large and sudden fluctuation of a finite-volume spectral density, that a covariance with vanishing correlation length such as the one in Eq.~\eqref{eq:hltgp_prior} can allow.

Let again the smearing kernel be the Gaussian $G_\sigma$. The prior of Eq.~\eqref{eq:hltgp_prior} leads to a posterior distribution for the spectral density smeared with $G_\sigma(\omega,E)$ that is centred around
\begin{equation}\label{eq:hltgp_central}
\rho^{\rm post}_\sigma(\omega) =  \rho^{\rm prior}_\sigma(\omega) + \sum_{\tau=1}^{\tau_{\rm max}}  g_\tau(\sigma, \omega) \, C(a \tau) \; ,
\end{equation}
with variance
\begin{equation}\label{eq:hltgp_variance}
\mathcal{K}^{\rm post}(\omega,\omega)  =  \left(\int dE \, G^2_\sigma(\omega,E) \, \frac{e^{\alpha \omega}}{\lambda} \right) - \sum_{\tau=1}^{\tau_{\rm max}}  g_\tau(\sigma, \omega) F^\sigma_\tau(\omega) \; .
\end{equation}
where the coefficients are exactly identical to the HLT ones. By additionally setting $\rho^{\rm prior}_\sigma(\omega)=0$ we therefore obtain the same answer of Ref.~\cite{Hansen:2019idp}. The only difference lies in the estimate of its error. In Ref.~\cite{Hansen:2019idp}, and Backus-Gilbert methods in general, the error is estimated from the available statistics, for instance with a bootstrap procedure. In the case of GPs, we have an analytic expression, Eq.~\eqref{eq:hltgp_variance}. Interestingly, the latter expression for the statistical error can be written as
\begin{equation}
\mathcal{K}^{\rm post}(\omega,\omega) = \int dE \,  \frac{e^{\alpha E}}{\lambda} \, G_\sigma(\omega,E) 
\left(  \sum_{\tau=1}^{\tau_{\rm max}} g_\tau(\sigma, \omega) \, b_T(a \tau,E) - G_\sigma(\omega,E) \right) \, ,
\end{equation}
which is reminiscent of the quantity used to monitor the systematic error in the context of Ref~\cite{Hansen:2019idp}. Indeed, both vanish in the limit of $\tau_{\rm max} \rightarrow \infty$.

\begin{figure}
    \centering
    \includegraphics[width=0.49\columnwidth]{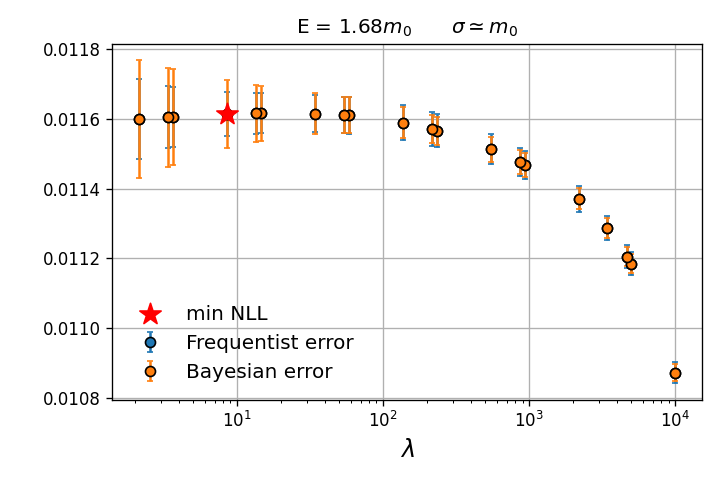}
    \includegraphics[width=0.49\columnwidth]{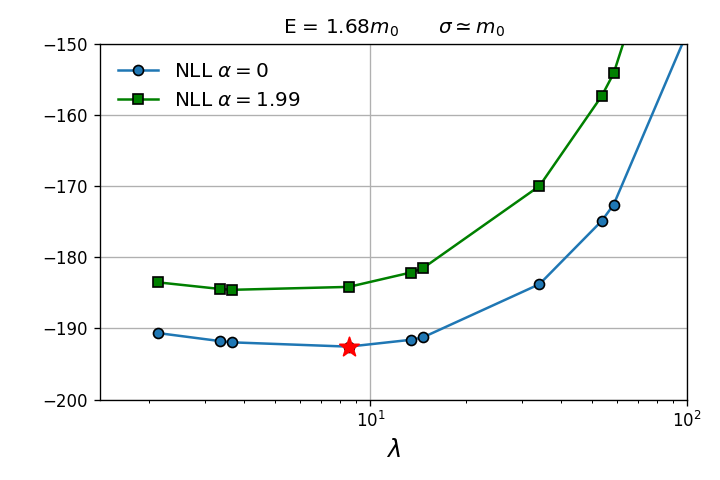}
    \caption{Left panel: spectral density smeared with a Gaussian, for different value of the $\lambda$-parameter. The central values are obtained according to Ref.~\cite{Hansen:2019idp}, or equivalently in the Bayesian setup of Section~\ref{sec:gphlt}. The error is computed with a bootstrap in the former case, and is given by Eq.~\eqref{eq:hltgp_variance} in the latter. The figure also shows the stability region described in Ref.~\cite{Bulava:2021fre}. In the right panel, we show corresponding values of the NLL around its minimum, including in this case $\alpha=1.99$. The starred point describes the minimum of the NLL (right) and the smeared spectral density at the value of $\lambda$ that gives such a value (left). The results are obtained using lattice data from Ref.~\cite{DelDebbio:2022qgu}.}
    \label{fig:gphlt-scan}
\end{figure}
The equivalence we just established allows to compare not only the results, but also certain systematics of the framework that lead to the same solution. In Fig.~\ref{fig:gphlt-scan} we show a comparison between the scan over the parameters established in Ref.~\cite{Bulava:2021fre} and the minimisation of the NLL. The latter, showed in the right panel, is minimised well within the stability region which we display on the left panel of the same figure. This compatibility suggests both procedures are solid. Two comments are however due: first, we observe that the Bayesian errors are generally larger. Moreover, while central values are the same for a fixed $\lambda$, an approach that is solely based on the stability region, could have picked a larger value of $\lambda$, since a plateau is reached before the NLL is minimised. These points will be clarified in our numerical experiments carried in the next section.

\section{Numerical tests on mock data}
\label{sec:toys}

In the previous sections, we established the close relation between a specific realisation of BG methods (Ref.~\cite{Hansen:2019idp}) and a specific realisation of Bayesian inference with GPs. Tests performed on lattice data in the previous sections suggest that the correspondence between these approaches may extend beyond the formal level, pertaining to practical applications. An important question that remains open is whether the procedures that are established here are able to produce unbiased results. A firm answer can hardly be given within this work, since data from lattice simulations can have a broad variety of features, and cases should be examined individually.

Here we propose to adopt a method to systematically validate a given setup for solving the inverse problem, by testing and tuning it against synthetic data generated according to a multivariate normal distribution characterised by the covariance matrix measured on the lattice, drawing inspiration from Ref.~\cite{DelDebbio:2021whr}. We will only compare the frequentist and Bayesian version of Ref.~\cite{Hansen:2019idp} here described in Sections~\ref{sec:hlt} and~\ref{sec:gphlt}, leaving out the ``traditional'' solution in terms of GPs from Section~\ref{sec:GP}: a comparison with the latter is still possible but less straightforward, since the smearing kernel is unconstrained.

We generated sets of $N_{\rm toys}$ correlators corresponding to a discrete spectrum of the following type:
\begin{equation}\label{eq:corr_rnd_peaks}
    C(t) = \sum_{n=0}^{n_{\rm max}-1} w_n \; e^{-|t|E_n} \; , \;\;\;\;\; E_0 < E_1 \leq \dots \; 
\end{equation}
The energies are taken in the interval [$2 m_\pi, 8 m_\pi$], where we set $m_\pi$ to be the mass of physical neutral pion. We use $n_{\rm max} \simeq 10$, which is a realistic number of energy levels for the typical lattice size of present simulations. The energy levels are taken as evenly spaced for simplicity. The weights $w_n$ are generated from a multivariate Gaussian distribution centred with vanishing average. The covariance has to reflect the distributional nature of the underlying finite-volume spectral density: we therefore choose
\begin{equation}\label{eq:toy_cov}
    K_{\rm weights}(n, n') = \kappa  \;  \exp \left( - \dfrac{(E_n-E_{n'})^2}{2 \epsilon^2} \right) \, ,
\end{equation}
with $\epsilon$ much smaller than the spacing between energy levels, and $\kappa$ characterising the variability of the weights.

\begin{figure}[htb]
    \centering
    \includegraphics[width=0.6\linewidth]{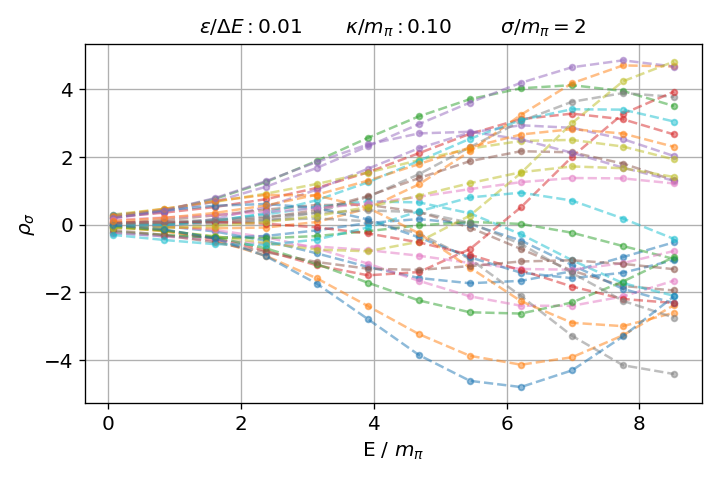}
    \caption{Instances of smeared spectral densities generated according to Eqs~\eqref{eq:corr_rnd_peaks} and Eq.~\eqref{eq:toy_cov}.}
    \label{fig:toys}
\end{figure}

In Fig.~\ref{fig:toys} we show several instances of weights transformed into smeared densities according to
\begin{equation}
    \rho_\sigma(E) = \sum_{n=0}^{n_{\rm max}-1} w_n \, G_\sigma(E-E_n) \; ,
\end{equation}
with smearing radius $\sigma=2 m_\pi$. For each of these sets of weights, we inject statistical noise into the corresponding correlator by using a covariance matrix computed in lattice QCD, for a vector-vector two-point correlation function of light-quark mesons.

We then solve the inverse problem on a large number of such toy correlators, looking for a statistically significant indication that one of the methods behave differently. As a measure of compatibility between the numerical prediction and the true result, we introduce the pull variable
\begin{equation}\label{eq:pull}
    p_\sigma(E) = \frac{\rho^{\rm pred}_\sigma(E) - \rho^{\rm true}_\sigma(E)}{\Delta \rho_\sigma} \, ,
\end{equation}
where $\rho^{\rm pred}_\sigma$ is the prediction for the smeared spectral density, $\rho^{\rm true}_\sigma$ is the exact solution, and the denominator is the total estimate of the uncertainty, statistical and systematic in quadrature. We shall also monitor the difference $\rho^{\rm pred}_\sigma(E) - \rho^{\rm true}_\sigma(E)$ without normalising it by the error.

\begin{figure}[htb]
    \centering
    \includegraphics[width=0.49\textwidth]{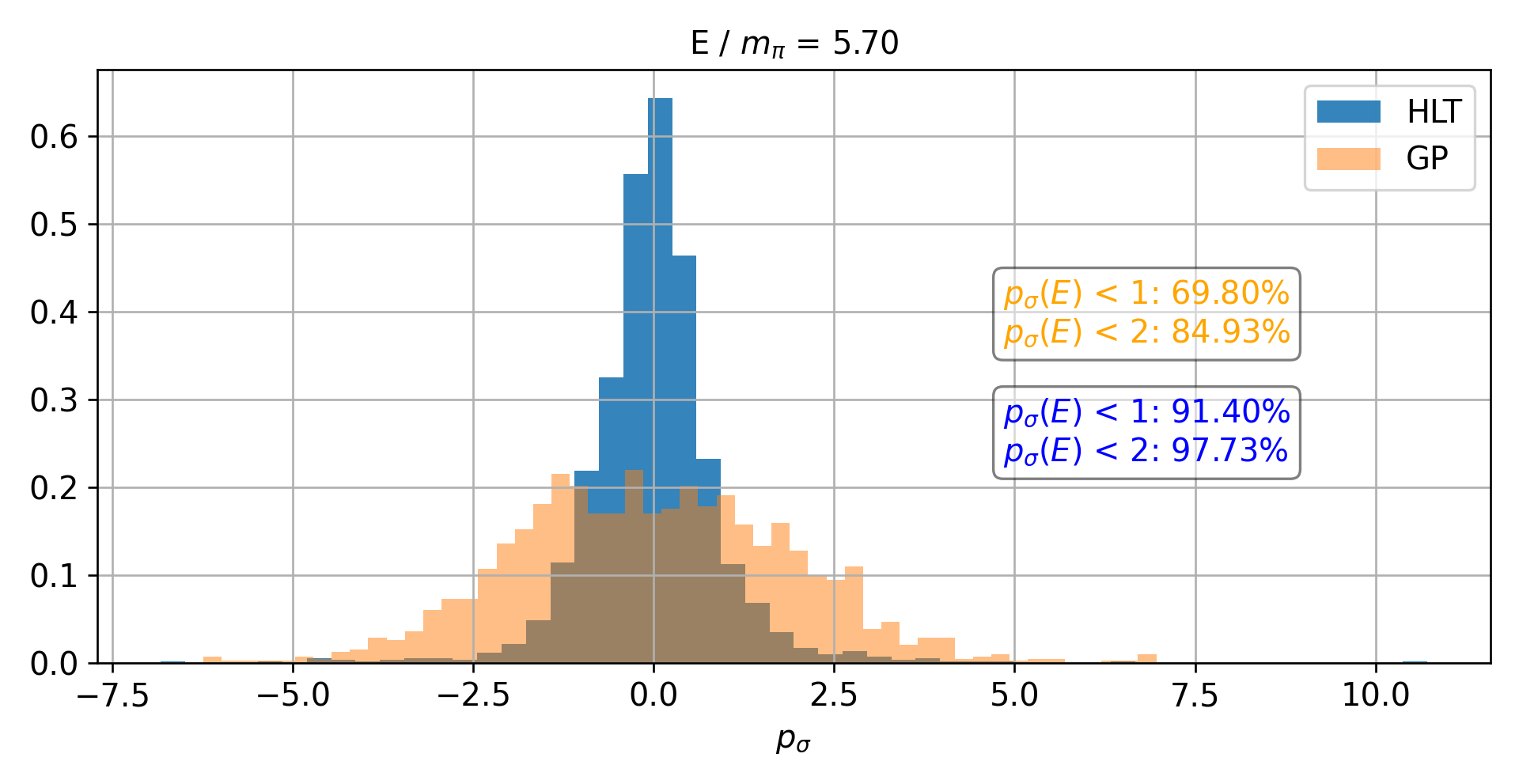}
    \includegraphics[width=0.49\textwidth]{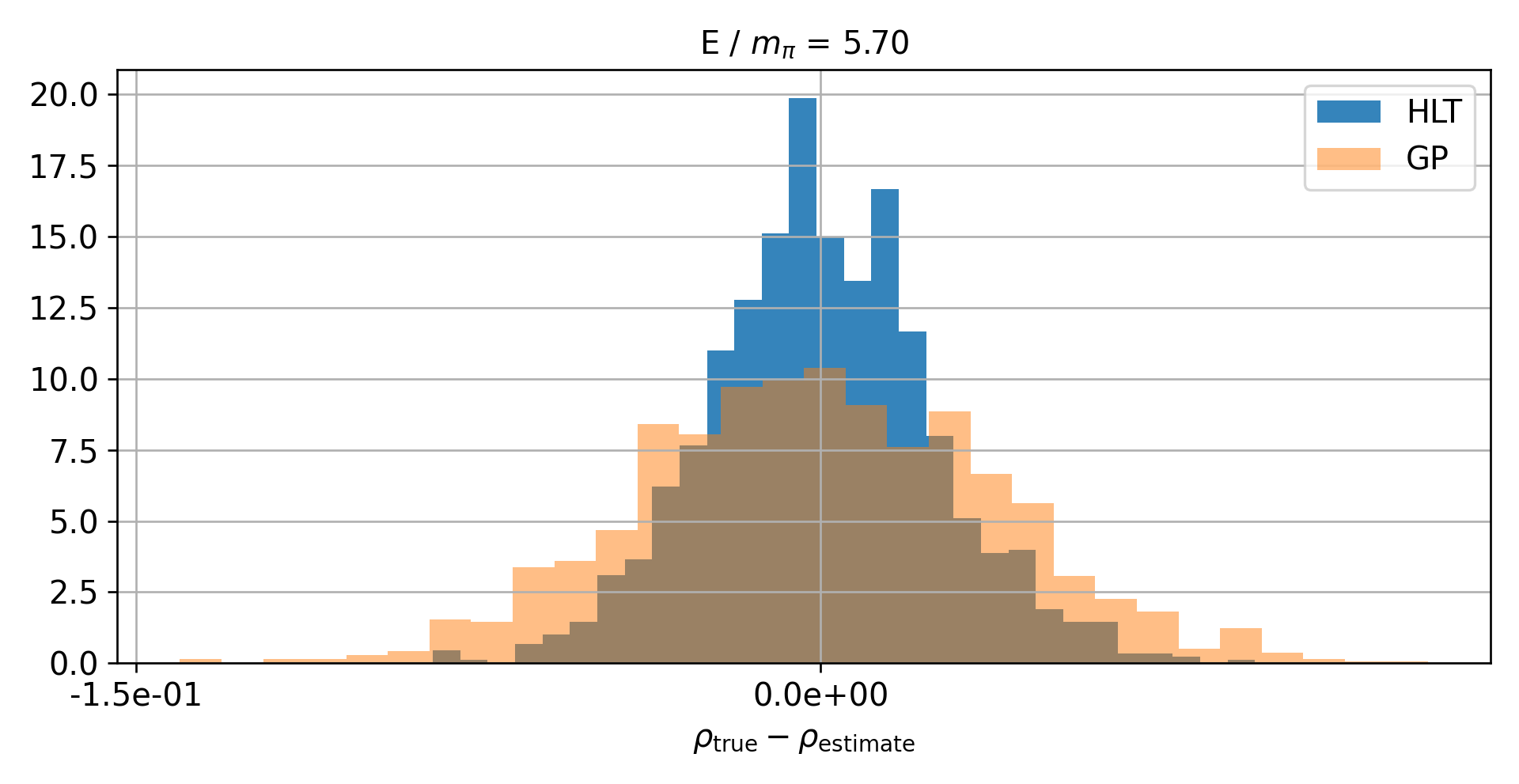}
    \caption{Histograms generated by solving the inverse problem for a thousand of instances of pseudo-data, as described in Section~\ref{sec:toys}. The left panel displays the pull variable defined in Eq.~\eqref{eq:pull}, using HLT (blue) and GP (orange). Additionally, the percentage of points within one and two standard deviations from the true result are shown. The difference in the two distributions are due to the different choice of $\lambda$ and the difference estimates of the error, as described in the text. In the right panel, the histogram shows the distribution of the difference between the prediction and the true smeared spectral density. The difference between the two distributions is, in this case, solely due to the different values of $\lambda$ that are prescribed.}
    \label{fig:histograms}
\end{figure}

\begin{figure}[htb]
    \centering
    \includegraphics[width=\linewidth]{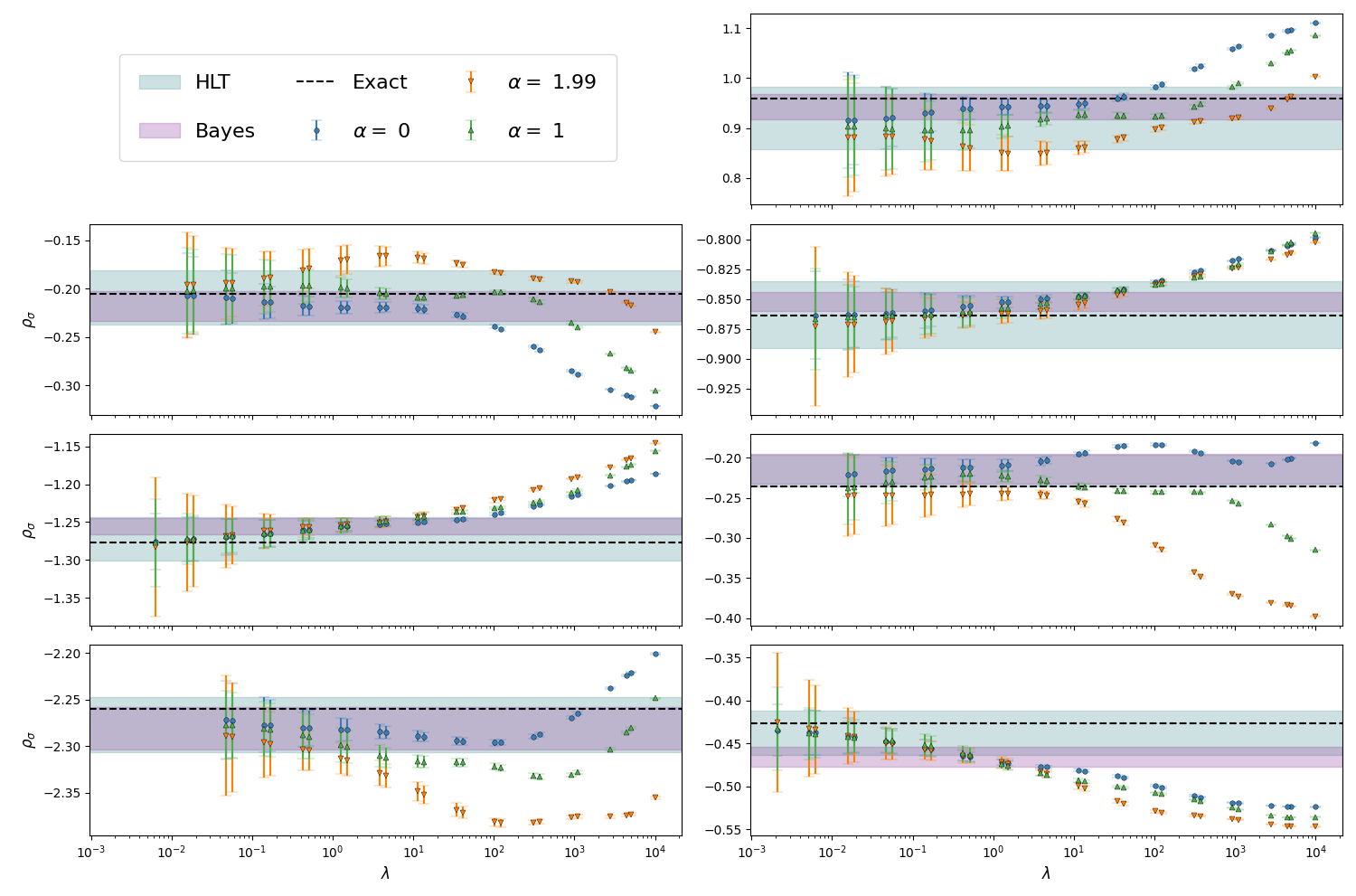}
    \caption{Examples of the stability analysis from the systematic analysis described in Section~\ref{sec:toys}. The HLT band is obtained by identifying a common plateau between different values of $\alpha$. The Bayesian result is obtained by minimising the NLL. The exact result is also shown as a black line. While this plot only shows seven instances, the results for a thousand samples are shown in Fig.~\ref{fig:histograms}.}
    \label{fig:lambda-scan-instances}
\end{figure}

We recall that only two sources of differences are possible given the way we set up the HLT and the GPs: the estimate of the error (obtained from resampling procedures in the former, and from Eq.~\eqref{eq:hltgp_variance} in the latter) and the value of $\lambda$ (plateau analysis in the former, analysis of the likelihood in the latter case). Fig.~\ref{fig:gphlt-scan} suggests that at fixed $\lambda$, the Bayesian error is more conservative. A more detailed picture can be however inferred from Fig.~\ref{fig:histograms}, showing the pull (cf. Eq.~\eqref{eq:pull}) on the left panel, and the non-normalised deviation from the true solution on the right. Differences in the latter plot can only derive from a difference selection of the parameter $\lambda$. The histogram corresponding to the HLT method is  narrower, meaning that in our working setup the plateau analysis is more effective in removing the effect of the bias, compared to the minimisation of the NLL. A moderate downside is that while the study of the NLL is easily automatised, the stability analysis requires more care, similarly to the fits of effective masses and correlation functions, where one has to scrutinise different fit ranges, correlations, etc. The plot on the left of the same figure shows that despite the difference in the errors at a fixed value of $\lambda$, the HLT method remains more conservative due to the smaller value of $\lambda$ that is chosen. The histograms in Fig.~\ref{fig:histograms} are the result of over 1000 instances of the inverse problems at the energy $E/ m_\pi = 5.7$, using $t_{\rm max}=32$ and a smearing radius $\sigma / m_\pi = 2$. In Fig.~\ref{fig:lambda-scan-instances} we show the scan over values of $\lambda$ corresponding to some of the cases (randomly chosen) entering the analysis. The horizontal bands are the prediction due to HLT and GPs. The former is obtained by identifying a plateau, the latter from the minimum of the NLL, as detailed in the previous sections. The exact solution is also displayed as a black horizontal line.

Finally, while the results shown in this section corresponding to a specific choice of energy, smearing radius, and other parameters, the limits in which the quality of the reconstruction improves or deteriorates are well understood, with better performances observed at smaller energies and larger values of $t_{\rm max}$ and $\sigma$ for either method.

\section{Conclusions}
In this work, we have reviewed two popular methods for regularising the inverse problem: the frequentist approach based on Ref.~\cite{Hansen:2019idp} which is close to a Backus-Gilbert solution, and a Bayesian approach based on Gaussian Processes~\cite{Horak:2021syv}. While both methods provide an answer in terms of a smeared spectral density, the latter does so, in its common implementation, by using an unconstrained smearing function (cf. Fig.~\ref{fig:naive_smearing_kernel}) which is not known a priori. We find this method to be therefore similar to the original Backus-Gilbert proposal. For the same reason, a systematic comparison of the results from Ref.~\cite{Hansen:2019idp} is not straightforward. We found, however, a striking similarity in the way the problem is regularised. In both cases, a matrix that is largely ill-conditioned due to the dumping exponential in the Laplace transform, is regularised with the covariance of the data. This step introduces a bias that is then treated differently in the two approaches: in the Bayesian case, one settles for the value of the regularising parameter that minimises the NLL. Applications of the HLT procedure, on the other hand, seek to absorb the dependence on this parameter into the statistical error by identifying a plateau in the small-$\lambda$ region.

In order to overcome the difficulties of an unconstrained smearing kernel, while at the same time retaining a Bayesian framework, we have shown in Section~\ref{sec:gphlt} how any Backus-Gilbert method, and in particular the one of Ref.~\cite{Hansen:2019idp}, can be formulated in a Bayesian way by setting up a probability density for a spectral function smeared with a chosen kernel, rather than one that is unsmeared. Together with the choice of a specific prior for the covariance of the smeared spectral densities that reproduces the correct distributional behaviour, this approach led to formulas that are in a one-to-one correspondence to those of Ref~\cite{Hansen:2019idp}. 

The only differences are formally the estimate of the error and the specific value of the regularising parameter $\lambda$. The error is found to be generally more conservative in the Bayesian case, at a specific value of $\lambda$. Concerning the choice of $\lambda$, which affects both the central value and the error, we have seen in Fig.~\ref{fig:gphlt-scan} examples in which the approaches are consistent. A more detailed analysis over pseudo-data generated according to a covariance matrix computed on the lattice suggests that the plateau analysis is more effective in removing the bias introduced by regularising the problem, as shown in Figs.~\ref{fig:histograms}, where the Bayesian methods produce results that are, on average, less reliable. While it is not clear whether this statement would hold for different varieties of  datasets, the analysis performed in Section~\ref{sec:toys} can be a powerful tool to assess the ability to produce unbiased predictions, and their usage is advocated. It is also worth noting that from a computational perspective, obtaining both the Bayesian and the frequentist solutions requires more or less the same amount of resources than working with just one, since the bottleneck of the inverse problem is the inversion of a matrix that is the same in both approaches. It could be therefore useful to tackle the inverse problem with this dual setup.

\section*{Acknowledgments}
A.L. is grateful to Ed Bennet and Niccolò Forzano for helping with the development of the code~\cite{Forzano:2024} used for this work. The authors express their sincere gratitude to Alessandro De Santis and the ETMC collaboration for providing the covariance matrix used to generate sets of mock data used in this work. We also thank Julian Urban for pointing to us important references as this work was being developed. 

A.L. is funded in part by l’Agence Nationale de la Recherche (ANR), under grant ANR-22-
CE31-0011. L.D.D. is funded by the UK Science and Tech-
nology Facility Council (STFC) grant ST/P000630/1
and by the ExaTEPP project EP/X01696X/1. M.P. has been partially supported by the Italian PRIN ``Progetti di Ricerca di Rilevante Interesse Nazionale -- Bando 2022'', prot. 2022TJFCYB, by the Spoke 1 ``FutureHPC \& BigData'' of the Italian Research Centre in High-Performance Computing, Big Data and Quantum Computing (ICSC), funded by the European Union -- NextGenerationEU, and by the SFT Scientific Initiative of the Italian Nuclear Physics Institute (INFN). N.T. is supported by the
Italian Ministry of University and Research (MUR) under the grant PNRR-M4C2-I1.1-PRIN 2022-
PE2 Non-perturbative aspects of fundamental interactions, in the Standard Model and beyond
F53D23001480006 funded by E.U. - NextGenerationEU.

\appendix
\section{Factorisation of the joint probability density}
\label{app:sec:LDU}
\subsection{With LDU decomposition}
Consider Gaussian probability density for the vector $\psi \in \mathbb{R}^{p+d}$ with covariance $\Sigma$,
\begin{equation}\label{eq:app:joint_unconditioned}
    \mathbb{G}[\psi;\Sigma] = \frac{1}{\sqrt{\det \left(2\pi \Sigma^{-1} \right)}} \exp \left( -\frac{1}{2} \,\psi^T \Sigma \, \psi \right) \; .
\end{equation}
With the example of Eq.~\eqref{eq:total_covariance} in mind, let the covariance be divided in blocks 
\begin{equation}
    \Sigma = \begin{pmatrix} \Sigma_{11} & \Sigma_{12} \\ \Sigma_{21} & \Sigma_{22} \end{pmatrix} \; , \;\;\;\;\; \psi = \begin{pmatrix} \phi_1 \\ \phi_2 \end{pmatrix} , \; 
\end{equation}
where $\phi_1 \in \mathbb{R}^{p}, \phi_2 \in \mathbb{R}^{d}$. $\Sigma_{11}$ is a $p \times p$ matrix, $\Sigma_{22}$ is $d\times d$, and $\Sigma_{12}$ and $\Sigma_{21}$ are $p\times d$ and $d \times p$ respectively. In order to get the conditioned probability density of $\phi_1$ given $\phi_2$, we can perform a LDU decomposition of the total covariance $\Sigma$. To this end, we introduce the matrices $L$ and $R$,
\begin{equation}
    L = \begin{pmatrix} \idd_p & - \Sigma_{12} \Sigma_{22}^{-1} \\ 0 & \idd_d \end{pmatrix} \; , \;\;\;\;\; R=\begin{pmatrix}  \idd_p & 0 \\ -\Sigma_{22}^{-1} \Sigma_{21} & \idd_d  \end{pmatrix} \; ,
\end{equation}
such that $W \equiv L \Sigma R$ is diagonal,
\begin{equation}
    W = \begin{pmatrix} \Sigma_{11} - \Sigma_{12} \Sigma_{22}^{-1} \Sigma_{21} & 0 \\ 0 & \Sigma_{22} \end{pmatrix} \; ,
\end{equation}
where the Schur complement of $\Sigma_{11}$ appears in the top-left block. The inverse of the covariance can be now written as
\begin{equation}
        \Sigma^{-1} = R \,  W^{-1} \,  L \; .
\end{equation}
The previous equations can be used to evaluate the scalar product $\psi^T \Sigma \psi$:
\begin{equation}
    \begin{split}
    & \psi^T \Sigma \, \psi = (\phi_1 - \phi_{(1|2)} )^T  \, \Sigma_{(11|2)}^{-1} \, (\phi_1 - \phi_{(1|2)} ) + \phi_2^T \, \Sigma_{22} \, \phi_2 \; , \\[8pt]
    & \phi_{(1|2)} \equiv \Sigma_{12} \Sigma_{22}^{-1} \phi_2 \; , \\[8pt]
    & \Sigma_{(11|2)} \equiv \Sigma_{11} - \Sigma_{12} \Sigma_{22}^{-1} \Sigma_{21} \;,
    \end{split}
\end{equation}
as well as the determinant of $\Sigma^{-1}$,
\begin{equation}
    \det \Sigma = \det\left(  \Sigma_{11} - \Sigma_{12} \Sigma_{22}^{-1} \Sigma_{21}\right) \det \left( \Sigma_{22} \right) \; . 
\end{equation}
As a consequence, the probability density of Eq.~\eqref{eq:app:joint_unconditioned} can be rewritten as
\begin{equation}\label{eq:app:joint_Conditioned}
    \mathbb{G}\left[\psi;\Sigma\right] = \mathbb{G}\left[\phi_1 - \phi_{(1|2)};\Sigma_{(11|2)}\right]  \mathbb{G}\left[\phi_2;\Sigma_{22}\right] \; .
\end{equation}
Factorising the conditional probability $\mathbb{G}\left[\phi_1 - \phi_{(1|2)};\Sigma_{(11|2)}\right]$ as in Eq.~\eqref{eq:conditional_probability_factorised}.

\subsection{In the Bayesian Language}
\label{app:sec:marginalise}
We can re-derive the expressions from the previous Section in the Bayesian language. Consider the probability density associated to the stochastic variable $\mathcal{R}(\omega)$ given prior value $\rho^{\rm prior}(\omega)$ and covariance $\mathcal{K}^{\rm prior}(\omega,E)$
\begin{equation}\label{eq:app:p_marginal_R}
    \pi\left(\mathcal{R}(\omega) \,| \; \rho^{\rm prior}(\omega), \,\mathcal{K}^{\rm prior}(\omega,E) \right) \; .
\end{equation}
The probability density associated to the variable $\hat{\vec{\mathcal{C}}}$ given the observed values $\vec{C}^{\rm obs}$ and covariance $\text{Cov}_d$ is similarly denoted by
\begin{equation}\label{eq:app:p_marginal_C}
    \pi \left( \hat{\vec{\mathcal{C}}} \, | \;  \vec{C}^{\rm obs} , \,\text{Cov}_d \ \right) \; .
\end{equation}
Notice that $\hat{\vec{\mathcal{C}}}$ is not the same as $\mathcal{C}$ introduced in Eq.~\eqref{eq:laplace_transform_with_noise}.                                           If the variables were to be independent, the joint probability would be the product of Eqs.~\eqref{eq:app:p_marginal_R} and~\eqref{eq:app:p_marginal_C}. In order to account for their correlation,
\begin{equation}
    \hat{\vec{\mathcal{C}}} = \vec{C}^{\rm obs} + \vec{\eta} \; .
\end{equation}
the joint probability becomes
\begin{multline}
    \pi \left( \mathcal{R}(\omega) , \, \hat{\vec{\mathcal{C}}} \,| \; \rho^{\rm prior}(\omega), \,\mathcal{K}^{\rm prior}(\omega,E), \, \vec{C}^{\rm obs} ,\, \text{Cov}_d  \right) \\= \pi\left(\mathcal{R}(\omega) \,| \; \rho^{\rm prior}(\omega), \,\mathcal{K}^{\rm prior}(\omega,E) \right) \, \pi \left( \hat{\vec{\mathcal{C}}} \, | \;  \vec{C}^{\rm obs} , \,\text{Cov}_d \ \right) N^{-1} \, \delta\left( \hat{\mathcal{C}}(t) - \int dE\; b_T(t,E) \, \mathcal{R}(E)  \right)\; ,
\end{multline}
where the factor $N$ ensures the proper normalisation for the joint probability after the addition of the delta function. We can then marginalise with respect to $\vec{\hat{\mathcal{C}}}$, obtaining
\begin{multline}\label{eq:app:marginalised_prob_density_GENERIC}
    \pi \left(  \mathcal{R}(\omega) \,| \; \rho^{\rm prior}(\omega), \,\mathcal{K}^{\rm prior}(\omega,E), \, \vec{C}^{\rm obs} ,\, \text{Cov}_d \right)\\ = N^{-1} \, \pi\left(\mathcal{R}(\omega) \,| \; \rho^{\rm prior}(\omega), \,\mathcal{K}^{\rm prior}(\omega,E) \right) \, \pi \left( \int dE\; b_T(t,E) \, \mathcal{R}(E) \, | \;  \vec{C}^{\rm obs} , \,\text{Cov}_d \ \right) \, .
\end{multline}
Consider the case, relevant for this work, in which all probability densities are Gaussian. Eq.~\eqref{eq:app:marginalised_prob_density_GENERIC} can be written as
\begin{multline}
 \pi \left(  \mathcal{R}(\omega) \,| \; \rho^{\rm prior}(\omega), \,\mathcal{K}^{\rm prior}(\omega,E), \, \vec{C}^{\rm obs} ,\, \text{Cov}_d \right)\\ =  \frac{N^{-1} }{\sqrt{\det \left(2\pi \text{Cov}_d \right)}}
    \exp \left( -\frac{1}{2} \left| \int dE_1 b_t(E_1) \mathcal{R}(E_1) - C^{\rm obs}(t) \right|_{\text{Cov}_{\rm d}}  \right) \\ \dfrac{1}{\sqrt{\det \left(2\pi  \mathcal{K}_{\rm prior} \right)}} \exp \biggr( -\frac{1}{2} \left| \mathcal{R} - \rho^{\rm prior} \right|^2_{\mathcal{K}^{\rm prior}} \biggr) \; ,
\end{multline}
which can be rewritten by completing the square in $\mathcal{R}$
\begin{multline}\label{eq:app:marginalised_prob_density_GAUSS}
     \pi \left(  \mathcal{R}(\omega) \,| \; \rho^{\rm prior}(\omega), \,\mathcal{K}^{\rm prior}(\omega,E), \, \vec{C}^{\rm obs} ,\, \text{Cov}_d \right) = \exp \left( \frac{1}{2} \left| q \right|^2_{\mathcal{K}^{\rm post}}    \right) \exp \left( -\frac{1}{2} \left| \rho^{\rm prior} \right|^2_{\mathcal{K}^{\rm prior}} \right) \\[6pt]
     \dfrac{1}{\sqrt{\det \left(2\pi  \mathcal{K}_{\rm prior} \right)}} \exp \left( -\frac{1}{2} \left| \mathcal{R} - \rho^{\rm post} \right|^2_{\mathcal{K}^{\rm post}} \right) 
      \frac{N^{-1} }{\sqrt{\det \left(2\pi \text{Cov}_d \right)}} \exp \left( -\frac{1}{2} \left| C^{\rm obs} \right|^2_{\text{Cov}_d} \right) \; ,
\end{multline}
where
\begin{align}
    & \rho^{\rm post}(\omega) = \int dE \, \left[ b_t(\omega) (\text{Cov}^{-1}_{\rm d})_{tr}  b_r(E) + \mathcal{K}^{-1}_{\rm prior}(\omega,E) \right]^{-1} \\ 
    & \hspace{5cm} \left[ b_{t'}(E)  (\text{Cov}^{-1}_{\rm d})_{t'r'} C^{\rm obs}_{r'} + \int dE' \, \mathcal{K}^{-1}_{\rm prior}(E,E') \rho^{\rm prior}(E') \right] \; ,\\[4pt]
    & \mathcal{K}^{\rm post}(\omega, E) = \left[ b_t(\omega)  (\text{Cov}^{-1}_{\rm d})_{tr}  b_r(E) + \mathcal{K}^{-1}_{\rm prior}(\omega,E) \right]^{-1} \; ,
\end{align}
and the exponential in $q(E)$ is the residual from completing the square in $\mathcal{R}(E)$:
\begin{equation}
    q(E) = \left[ b_{t}(E)  (\text{Cov}^{-1}_{\rm d})_{tr} C^{\rm obs}_{r} + \int dE' \, \mathcal{K}^{-1}_{\rm prior}(E,E') \rho^{\rm prior}(E') \right] \; .
\end{equation}
It is possible to rewrite the covariance of the posterior probability density by generalising the Woodbury identity,
\begin{equation}
    \mathcal{K}^{\rm post}(\omega, E) = \mathcal{K}^{\rm prior}(\omega, E) - F_t(\omega) (\text{Cov}_{\rm d} + \Sigma)^{-1}_{tr} F_r(E)\; ,
\end{equation}
and consequently its central value,
\begin{equation}
    \rho^{\rm post}(\omega) = \rho^{\rm prior}(\omega) + F_t(\omega) (\text{Cov}_{\rm d} + \Sigma)^{-1}_{tr} \left[ C^{\rm obs}(r) - C^{\rm prior}(r) \right] \; ,
\end{equation}
where we defined, in compliance with the main body of this work,
\begin{align}
    & \Sigma_{tr} = \int dE_1 dE_2 \, b_t(E_1) \mathcal{K}^{\rm prior}(E_1,E_2) b_r(E_2) \\
    & F_t(\omega) = \int dE_1 \, \mathcal{K}^{\rm prior}(\omega,E_1) b_t(E_1) \; , \\
    & C^{\rm prior}(t) = \int dE \, \rho(E) b_t(E) \; .
\end{align}
Finally, the data likelihood can be obtained from Eq.~\eqref{eq:app:marginalised_prob_density_GENERIC} by completing the square in $\vec{C}^{\rm obs}$: 
\begin{multline}
    \pi \left(  \mathcal{R}(\omega) \,| \; \rho^{\rm prior}(\omega), \,\mathcal{K}^{\rm prior}(\omega,E), \, \vec{C}^{\rm obs} ,\, \text{Cov}_d \right) = \dfrac{1}{\sqrt{\det \left(2\pi  \mathcal{K}_{\rm post} \right) \det \left(2\pi (\Sigma + \text{Cov}_{\rm d}) \right)}}   \\[4pt]
    \exp \left( -\frac{1}{2}\left| \mathcal{R}-\rho^{\rm post} \right|^2_{\mathcal{K}_{\rm post}}  \right) \, \exp \left( -\frac{1}{2} \left| \vec{C}^{\rm obs} - \vec{C}^{\rm prior} \right|^2_{\Sigma + \text{Cov}_{\rm d}} \right) \; ,
\end{multline}
where we have used $N = \sqrt{\frac{\det \left(2\pi  \mathcal{K}_{\rm post} \right) \det \left(2\pi (\Sigma + \text{Cov}_{\rm d} )\right)}{ \det \left(2\pi  \mathcal{K}_{\rm prior} \right) \det (2\pi \text{Cov}_{\rm d})}}$.

\newpage
\bibliographystyle{JHEP}       
\bibliography{main}

\end{document}